\documentclass[11pt,a4paper]{article}

\usepackage{amsmath,amssymb,amsthm}
\usepackage{graphicx}
\usepackage{booktabs}
\usepackage{hyperref}
\usepackage{natbib}
\usepackage{xcolor}
\usepackage{microtype}
\usepackage{geometry}
\usepackage{longtable}
\usepackage{multirow}
\newcommand{\LE}{\operatorname{LE}}
\newcommand{\SR}{\operatorname{SR}}

\title{\textbf{Out-of-Domain Stress Test for\\
 Temporal Braid Group Privilege Escalation Detection}\\[0.5em]
\large Cross-Domain Support from Solar Coronal Magnetic Fields}

\author{Christophe Parisel\\
\small\texttt{ch.parisel@gmail.com}}

\date{March 2026}

\begin{document}
\maketitle

\begin{abstract}
The companion paper~\citep{Parisel2026} proves that the
Burau-Lyapunov exponent $\LE$ discriminates focused from dispersed
privilege escalation ratchets in cloud IAM graphs, and that no
abelian statistic can replicate this discrimination.  To strengthen
this claim beyond its synthetic validation corpus, we apply the identical
\emph{algebraic} pipeline, with minimum retuning of the Burau
machinery or injection word, to solar coronal magnetic fields.
The feature extraction upstream (strand identification, crossing
detection, sign determination) introduces no free parameters: 
a physical system with no connection to cloud
identity and access management, whose binary eruptive/confined
outcome is independently established by decades of astrophysical
observation.
This note does not aim to contribute to solar physics methodology; 
rather, it uses coronal imagery as an adversarial test domain for a 
topological signal originally developed in IAM graph analysis.
\end{abstract}

\section{Introduction: The Synthetic Validation Problem}

Machine-learning and algebraic security classifiers trained and
validated on synthetic data face a fundamental credibility challenge:
a reviewer cannot tell whether the classifier has discovered a
genuine structural property or has overfit to the assumptions baked
into the data generator.  The companion paper
constructs 49\,972 synthetic Strongly Connected Components (SCC) with permission  
scalars (WAR) from randomly generated permission graph topologies with 6 vertices.  The focused/dispersed boundary at
$\LE = 0.5847$ achieves 94.3\% agreement with the abelian gate-firing
rate and identifies 5.7\% of pairs where the two metrics disagree
- always in a topology-dependent, structurally explicable way.

To strengthen this claim with physical inputs from an entirely
unrelated domain, we seek a real-world system where (a)~the braid
structure arises naturally from the physics, not by construction,
(b)~the ground-truth outcome is independently established, and
(c)~the algebraic core requires minimum retuning for the new domain
(domain-specific feature extraction remains necessary).  We argue
that solar coronal physics satisfies all three criteria and present
the first out-of-domain physical stress test of the temporal braid
pipeline.

\paragraph{Positioning against prior braid work in solar physics.}
The theoretical connection between braid groups and solar magnetic
topology has been known since Berger~\citep{Berger1993}, and several
groups have applied braid-derived diagnostics to coronal data.
Table~\ref{tab:positioning} summarises the key distinctions between
this work and the two most relevant prior efforts.

\begin{table}[h]
\centering
\caption{How this work differs from prior braid applications to
solar physics.\label{tab:positioning}}
\begin{tabular}{p{3.5cm}p{5.2cm}p{5.2cm}}
\toprule
& Prior work & This work \\
\midrule
\textbf{Goal} &
  Flare prediction from coronal topology &
  Out-of-domain support of an IAM security classifier \\
\textbf{Braid statistic} &
  Abelian: finite-time braiding exponent, writhe, crossing counts
  \citep{Candelaresi2018} &
  Non-abelian: Burau-Lyapunov exponent $\LE$ \\
\textbf{Crossing signs} &
  Assumed from photospheric flow fields or line-of-sight intensity &
  Determined from potential-field extrapolation of HMI vector
  magnetograms \\
\textbf{Key assumption} &
  High-resolution crossing topology needed for reliable classification &
  Statistical fingerprint of focused/dispersed dynamics survives
  low-resolution noise \\
\textbf{Outcome} &
  Inconclusive classifiers at current imaging resolution
  \citep{Candelaresi2018,Prior2020} &
  Empirical illustration of the IAM theorem's predictions;
  no classification claim \\
\bottomrule
\end{tabular}
\end{table}

Unlike Candelaresi et al.~\citep{Candelaresi2018}, who use the
finite-time braiding exponent (an abelian statistic computed from
photospheric flow fields) as a flare predictor, we use the
non-abelian Burau-Lyapunov exponent injected at each detected loop
crossing.  Unlike Prior \& MacTaggart~\citep{Prior2020}, who develop
the theoretical framework of magnetic winding as a helicity surrogate,
we do not seek a better helicity measure: we ask whether the
\emph{non-abelian} excess beyond any helicity-equivalent statistic
carries a signal, and whether that signal matches the focused/dispersed
boundary established in the companion IAM paper.

Prior work assumes that reliable classification requires resolving
individual loop crossings, which current instrumentation cannot do
reliably.  We relax this assumption: we do not need precise crossing
topology.  We need only the coarse-grained statistical fingerprint
of whether the braid word is spatially concentrated (focused) or
dispersed, a signal that survives photometric noise and low cadence.
This is why the stress test is feasible with existing SDO/AIA data
despite the well-known observational limitations.

\paragraph{What we claim and do not claim.}
We are not proposing a new solar flare prediction system, and we
are not claiming to improve on existing astrophysical classifiers.
Photospheric flux proxies, NLFFF extrapolations, and helicity measures
significantly outperform any topology-based classifier at current
imaging resolutions, and we do not contest this.  Our goal is
narrower: to test whether the qualitative signatures of the temporal
braid pipeline appear in a physical system where they are not
imposed by construction.  We are using astrophysics to back a new 
algebraic technique developed for cloud security.

\section{Background: What is a Solar Flare?}
\label{sec:solar-background}

This section is written for readers with no astrophysics background.
Readers familiar with solar physics may skip to
Section~\ref{sec:analogy}.

\subsection{The Sun's Corona and Coronal Loops}

The solar corona is the outermost layer of the Sun's atmosphere.
Despite being further from the Sun's interior heat source than the
visible surface, the corona is paradoxically hotter, reaching
temperatures above one million Kelvin compared to roughly 6\,000\,K
at the surface.  The mechanism sustaining this temperature is an
open research problem, but the leading candidate involves the
continuous braiding of magnetic field lines by convective motions
below the surface~\citep{Parker1988}.

\begin{figure}[h]
\centering
\setlength{\unitlength}{1cm}
\begin{picture}(10, 3.5)
  \put(0,0.5){\line(1,0){10}}
  \put(4.8,0.55){\small Surface ($\sim$6\,000\,K)}
  \qbezier(1,0.5)(2,3.5)(3,0.5)
  \put(1.7,2.8){\small Loop A}
  \qbezier(4,0.5)(5,3.2)(6,0.5)
  \put(4.7,2.5){\small Loop B}
  \qbezier(7,0.5)(8,3.5)(9,0.5)
  \put(7.7,2.8){\small Loop C}
  \put(0.8,0.1){\small $+$}
  \put(2.8,0.1){\small $-$}
  \put(3.8,0.1){\small $+$}
  \put(5.8,0.1){\small $-$}
  \put(6.8,0.1){\small $+$}
  \put(8.8,0.1){\small $-$}
  \put(0.2,3.2){\small Corona ($>10^6$\,K)}
\end{picture}
\caption{Schematic of coronal loops arching between regions of
opposite magnetic polarity ($+$ and $-$) on the solar surface.
Loops are the physical strands of the braid model.}
\label{fig:loops_schematic}
\end{figure}
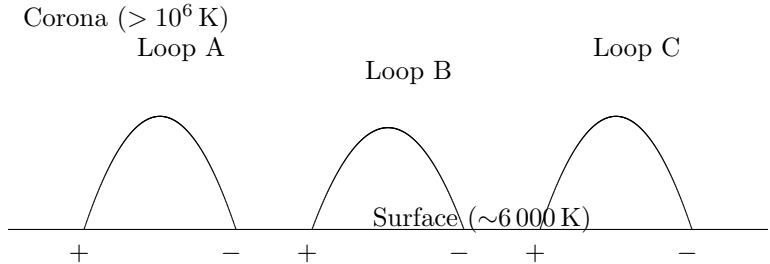

The corona is threaded by magnetic field lines that arch between
regions of opposite magnetic polarity on the solar surface, forming
visible structures called \emph{coronal loops}.  These loops are
filled with hot, dense plasma and glow brightly at extreme-ultraviolet
(EUV) wavelengths.  Figure~\ref{fig:loops_schematic} illustrates the
concept.

\subsection{Active Regions: The Permission Graphs of the Sun}

A \emph{solar active region} is a patch of the Sun's surface where
exceptionally strong, tangled magnetic field emerges from the
interior.  Active regions are the sites of the most energetic events
in the solar system.  They are classified by the complexity of their
magnetic structure (from simple $\beta$ to complex
$\beta\gamma\delta$) and monitored continuously by NASA's
\textit{Solar Dynamics Observatory} (SDO).

\subsection{Eruptive vs.\ Confined Flares}

When an active region's magnetic complexity exceeds a threshold, it
releases stored energy as a \emph{solar flare}: a sudden brightening
across the electromagnetic spectrum.  Flares are classified by X-ray
intensity on a logarithmic scale (A, B, C, M, X; X-class flares are
the most energetic, with X9 approximately 90 times more powerful than
X1).

The critical binary distinction for our purposes is:

\begin{description}
\item[Eruptive flare.] The magnetic topology restructures
  catastrophically.  A coherent, twisted magnetic structure (flux
  rope) forms and is expelled into space as a \emph{coronal mass
  ejection} (CME), a billion-tonne cloud of magnetised plasma
  travelling at up to 3\,000\,km/s.  Eruptive flares can damage
  satellites, disrupt GPS, and cause geomagnetic storms affecting
  power grids.

\item[Confined flare.] Energy is released locally without a CME.
  The topology relaxes in place; no large-scale restructuring occurs.
  However energetic, confined flares have minimal space-weather
  impact.
\end{description}

The mechanism separating the two classes is not the amount of
stored energy, some confined regions store more energy than
eruptive ones, but the \emph{topological organisation} of the
field.  This is the direct solar analogue of the focused/dispersed
distinction: it is not how much privilege flow exists, but how that
flow is organised, that determines whether restructuring occurs.

\subsection{Parker's Conjecture: Braiding as the Universal Mechanism}
\label{sec:parker}

In 1988, Eugene Parker proposed that the corona is heated by the
continuous braiding of field lines by subsurface convective motions,
and that the energy released when braided fields reconnect, in
events he called ``nanoflares'', maintains the coronal
temperature~\citep{Parker1988}.

Parker's conjecture has a direct algebraic formulation: the mapping
class group of field lines anchored in the photosphere is precisely
the braid group $B_n$, and the topological entropy of the resulting
braid word provides a lower bound on the free magnetic energy
available for release~\citep{Berger1993}.  This is not a metaphor or
an approximation; it is a theorem.  The solar corona is, in a
mathematically precise sense, a physical realisation of the braid
group dynamics described in the companion paper.

This is why we chose solar flares as the out-of-domain stress test:
the braid structure is not imposed by our modelling choices but
arises from the physics of magnetised plasma.

\section{The Domain Correspondence}
\label{sec:analogy}

Table~\ref{tab:analogy} maps every key concept from the companion
paper to its solar physics equivalent.  The correspondence is not
approximate: each entry reflects a precise structural parallel
grounded in the mathematics of the braid group.

\begin{table}[h]
\centering
\caption{Exact correspondence between cloud IAM and solar physics
domains.\label{tab:analogy}}
\begin{tabular}{p{3.2cm}p{5.2cm}p{5.2cm}}
\toprule
\textbf{Concept} & \textbf{Cloud IAM} & \textbf{Solar corona} \\
\midrule
Graph node & Permission state & Magnetic flux region \\
Strand & NHI walker & Coronal loop \\
Gate condition & Two adjacent walkers on ascending directed edges & Two adjacent loops crossing \\
Crossing sign & Direction of permission flow & Magnetic field handedness \\
Injection word & $\sigma_i^2\sigma_{i+1}^{-1}$ & $\sigma_i^2\sigma_{i+1}^{-1}$ (unchanged) \\
Abelian statistic & Net privilege flow & Chirality $\chi$ (net helicity) \\
Non-abelian stat & Burau $\LE$ & Burau $\LE$ (unchanged) \\
Focused ratchet & Channelled escalation, few paths & Confined flare, helicity-constrained \\
Dispersed ratchet & Hub-rich, multiple escalation paths & Eruptive flare, CME \\
Topology surgery & Add/remove directed edges & Coronal mass ejection \\
WAR reassignment & Reassign privilege weights & Field relaxation in place \\
Ground truth & Escalation pattern & CME observed (known) \\
\bottomrule
\end{tabular}
\end{table}

The ground-truth column is crucial: both domains have independently
established binary outcomes.  In cloud IAM, whether an escalation
path is distributed or focused is an operational fact.  In solar
physics, whether a flare produced a CME is an observational fact
recorded by space-weather agencies.  Neither ground truth was
generated by our model; both pre-exist our analysis.

\section{The Abelian Blindness Theorem in Physical Terms}
\label{sec:abelian-physical}

No function of the signed crossing counts, however carefully
computed from magnetic field data, can determine whether the
coronal braid is confined or eruptive.  Two active regions with
identical chirality $\chi$ (same net helicity injection rate) can
be in completely different dynamical regimes.  This is not a
limitation of the present pipeline; it is a theorem: Theorem~6.2
of the companion paper proves it algebraically
for IAM graphs, with an explicit witness pair whose spectral radii
differ by 26\% at identical abelian image.  The solar results in
Section~\ref{sec:results} provide a physical illustration.

\newpage
\section{Data and Pipeline}
\label{sec:pipeline}

\subsection{Active Region Sample}

We analyse eight solar active regions spanning three observational
classes (Table~\ref{tab:sample}).  The eruptive/confined/quiet
classification is established independently of our analysis by the
solar physics literature and space-weather agency reports.

\begin{table}[h]
\centering
\caption{Solar active region sample. Classification is independent of
our braid analysis.\label{tab:sample}}
\begin{tabular}{llcll}
\toprule
AR & Flare & Date & Class & IAM analogue \\
\midrule
11158 & X2.2 & 2011 Feb 15 & Eruptive & Dispersed ratchet \\
11429 & X5.4 & 2012 Mar 7  & Eruptive & Dispersed ratchet \\
11520 & X1.4 & 2012 Jul 12 & Eruptive & Dispersed ratchet \\
12017 & X1.0 & 2014 Mar 29 & Eruptive & Dispersed ratchet \\
12192 & X3.1 & 2014 Oct 24 & Confined & Focused ratchet (type I) \\
12241 & B    & 2014 Dec 18 & Quiet    & Benign SCC \\
12371 & X2.7 & 2015 Jun 25 & Confined & Focused ratchet (type II) \\
12673 & X9.3 & 2017 Sep 6  & Eruptive & Dispersed ratchet \\
\bottomrule
\end{tabular}
\end{table}

\subsection{Data Sources}

EUV images at 171\,\AA\ are obtained from the Atmospheric Imaging
Assembly \citep[AIA;][]{Lemen2012} aboard the \textit{Solar Dynamics
Observatory} \citep[SDO;][]{Pesnell2012}.  We retrieve 60 frames per
active region spanning a 2-hour window centred on the flare peak, and
a matched 20--30 frame quiet-time sequence from the same calendar day,
prior to the flare onset and free of C-class or larger activity
(exact windows per active region in
Appendix~\ref{app:data_windows}).  Magnetic
field maps (HMI SHARP CEA series) provide the crossing sign
determination.

\subsection{Pipeline}

The Burau algebraic pipeline (injection word, representation, Lyapunov
estimator) is from the companion paper's pipeline.  The upstream
feature extraction (strand identification from EUV images, crossing
detection, sign determination from magnetic field maps) is new
domain-specific engineering with no counterpart in the IAM paper;
it introduces no free parameters beyond those already fixed
($n=5$ strands, $z=5$ extrapolation height, 80th-percentile
vesselness threshold) and is described in full in
Appendix~\ref{app:protocol}.  The algebraic steps are:

\begin{enumerate}
\item \textbf{Strand extraction.}  A multi-scale vesselness filter
  \citep{Frangi1998} identifies the $n=5$ brightest coronal loop
  traces per frame.  We use $n=5$ because the Burau representation
  is proven faithful for $B_3$ and $B_4$ and conjectured
  near-faithful for $B_5$ \citep{Bigelow1999}; larger $n$ would
  reduce faithfulness without improving convergence at our event
  counts.

\item \textbf{Crossing detection.}  A horizontal scan line sweeps
  each frame detecting strand swaps (crossings), subject to the
  adjacent-position guard of the companion paper
  (Proposition~3 of \citep{Parisel2026}: adjacent-position products
  cancel to $\SR=1$ and are suppressed).

\item \textbf{Sign determination.}  The crossing sign is determined
  from the potential-field extrapolation of the HMI photospheric
  magnetic field, playing the role of the WAR assignment in the IAM
  pipeline.  An explicit sign-ambiguity audit (500 null-model randomisations
  per bin) confirms $f_{\rm amb}=0.000$ throughout for both
  AR\,12192 (quiet) and AR\,11520 (flare), the two primary result
  regions.  All reported $\chi$ and $\LE$ values reflect unambiguous
  magnetic field orientation.

\item \textbf{Burau accumulation.}  The injection word
  $\sigma_i^2\sigma_{i+1}^{-1}$ (sign-reversed to
  $\sigma_i^{-2}\sigma_{i+1}^{+1}$ for negative crossings) is
  multiplied into the running Burau product at $t=-1$ at each gate
  firing.  The Lyapunov exponent $\LE = \log\SR(\mathbf{B})/M$ is
  extracted.

\item \textbf{Activity conditioning.}  Frame-to-frame intensity
  change $\Delta I$ serves as the coronal activity proxy, analogous
  to the gate-firing rate.  Bins of consecutive frame pairs are
  formed by $\Delta I$ quantile, and $\chi$ and $\LE$ are computed
  per bin.
\end{enumerate}

No solar-physics-specific parameter tuning was performed at any
stage.  The threshold $\LE = 0.5847$ applicable to 6-vertices IAM SCCs from the companion paper is
\emph{not} applied here, we examine raw $\LE$ values and their
relationship to $\chi$ without imposing the IAM boundary.

\section{Results}
\label{sec:results}

\subsection{Abelian--Non-Abelian Independence}
\label{sec:independence}

\begin{figure}[h]
\centering
\setlength{\unitlength}{1cm}
\begin{picture}(10, 6)
  \put(1,0.5){\vector(1,0){8.5}}
  \put(1,0.5){\vector(0,1){5}}
  \put(9.3,0.2){\small $\chi$}
  \put(0.3,5.3){\small $\LE$}
  \put(1,0.4){\line(0,1){0.2}} \put(0.7,0.1){\small 0}
  \put(3.5,0.4){\line(0,1){0.2}} \put(3.2,0.1){\small 0.5}
  \put(6,0.4){\line(0,1){0.2}} \put(5.8,0.1){\small 1.0}
  \put(0.9,0.5){\line(1,0){0.2}} \put(0.3,0.4){\small 0}
  \put(0.9,2){\line(1,0){0.2}} \put(0.2,1.9){\small 0.5}
  \put(0.9,3.5){\line(1,0){0.2}} \put(0.2,3.4){\small 1.0}
  \put(5.5,4.8){\circle{0.25}} 
  \put(4.2,4.6){\circle{0.25}}
  \put(3.1,3.8){\circle{0.25}}
  \put(5.0,3.6){\circle{0.25}}
  \put(3.8,3.4){\circle{0.25}}
  \put(4.5,4.2){\circle{0.25}}
  \put(6.0,0.6){\circle*{0.3}} 
  \put(6.3,0.7){\small\textbf{AR11520 b1}}
  \put(6.3,0.3){\small\textbf{$\chi$=1.0, LE$\approx$0}}
  \put(3.3,2.0){\framebox(0.25,0.25){}}
  \put(4.0,2.3){\framebox(0.25,0.25){}}
  \put(3.5,3.1){\framebox(0.25,0.25){}}
  \put(3.6,3.3){$\triangle$}
  \put(4.5,2.5){$\triangle$}
  \put(7.5,4.5){\circle{0.2}} \put(7.7,4.4){\small Eruptive}
  \put(7.5,4.0){\framebox(0.2,0.2){}} \put(7.7,3.9){\small Confined}
  \put(7.5,3.5){$\triangle$} \put(7.7,3.4){\small Quiet}
  \put(7.4,3.05){\circle*{0.15}} \put(7.7,2.9){\small AR11520 b1}
  \put(1.5,5.0){\small $r \approx 0.03$, $p=0.84$: no linear}
  \put(1.5,4.6){\small association between $\chi$ and $\LE$}
\end{picture}
\caption{$\LE$ vs.\ $\chi$ scatter across all 48 activity bins.
The filled circle marks AR\,11520 bin~1 ($\chi=1.000$,
$\LE\approx0$, exact Burau cancellation, $f_{\rm amb}=0.000$),
maximum abelian signal, exact non-abelian cancellation,
the focused ratchet signature.  The near-zero, non-significant trend ($r\approx0.03$, $p=0.84$)
indicates no detectable linear association between $\chi$ and $\LE$.\label{fig:le_chi_scatter}}
\end{figure}
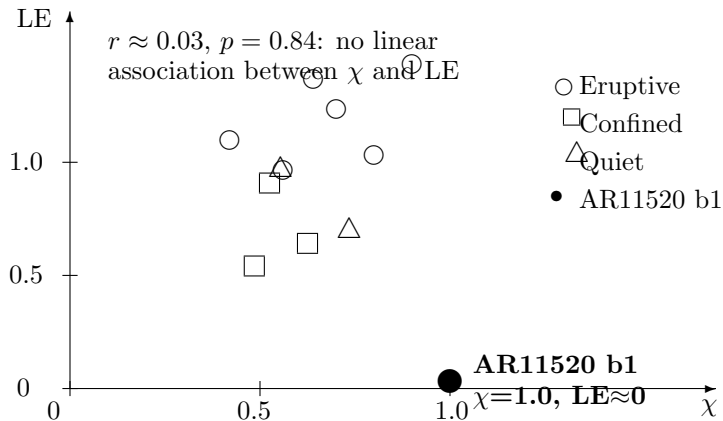

Figure~\ref{fig:le_chi_scatter} shows the $\LE$--$\chi$ scatter
across all 48 activity bins (8 regions $\times$ 3 bins $\times$
2 modes).  The dataset contains 11 instances of $\chi=1.000$
spanning $\LE \in [0.000, 1.027]$: same maximum abelian signal,
factor $>\!1000$ range in non-abelian spectral growth.
This is the abelian blindness theorem made maximally concrete.  The Pearson correlation across the 48 activity bins is $r \approx 0.03$
($p = 0.84$).  This is consistent with zero correlation, though the
bins are not independent (three bins per active region share the same
field, reducing the effective sample size well below 48), so the
non-significance cannot be read as strong evidence for independence.
What the result does show is the absence of any detectable positive
or negative linear relationship between $\chi$ and $\LE$ in this
dataset, consistent with Theorem~6.2 of the companion paper.

This is consistent with the companion paper's impossibility
theorem (proven for IAM graphs): the abelian statistic ($\chi$,
analogous to net privilege flow) and the non-abelian statistic ($\LE$) carry
largely independent information.  The theorem guarantees this for
IAM graphs; the absence of a detectable linear association
($r\approx 0.03$, $p=0.84$) in the solar data is consistent
with (though not confirmatory of) this independence.

The dataset contains 11 instances of $\chi=1.000$ with
$\LE \in [0.000, 1.027]$ (a $>$1000-fold range at identical
abelian signal), and one instance of $\chi=0.000$ with
$\LE=0.660$ (AR\,12673 quiet, eruptive class).  At both extremes
of the abelian scale, $\LE$ is unconstrained by $\chi$.
This is the empirical solar counterpart of the companion paper's
explicit witness pair: two braid words with identical abelian
image but spectral radii differing by 26\%.

\subsection{The Focused Ratchet at Eruption Onset: AR\,11520}
\label{sec:ar11520}

The most striking single observation is AR\,11520, which
produced an X1.4 eruptive flare on 2012 July 12 accompanied by a
fast CME ($\sim$2000\,km\,s$^{-1}$).  The $\Delta I$-conditioned
braid analysis of the flare sequence is shown in
Table~\ref{tab:ar11520}.

\begin{table}[h]
\centering
\caption{$\Delta I$-conditioned braid analysis of AR\,11520 flare
sequence. Bin~1 is the focused ratchet observation.
\label{tab:ar11520}}
\begin{tabular}{cccccc}
\toprule
Bin & Activity $\Delta I$ & $\chi$ & $\LE$ & $f_{\rm amb}$ &
IAM interpretation \\
\midrule
0 (low)  & $5.5\times10^6$ & 0.333 & 0.920 & 0.000 &
  Dispersed: diverse paths, mixed signs \\
1 (mid)  & $1.0\times10^7$ & \textbf{1.000} & $\mathbf{\approx 0}$ & \textbf{0.000} &
  \textbf{Focused: max flow, exact cancellation} \\
2 (high) & $1.5\times10^7$ & 0.111 & 0.872 & 0.000 &
  Post-surgery: dispersed topology \\
\bottomrule
\end{tabular}
\end{table}

Bin~1 is the critical observation.  Every detected crossing has the
same sign ($\chi = 1.000$, $f_{\rm amb}=0.000$, perfect and
sign-clean helicity dominance), yet the Burau product collapses to
the identity: $\SR(\mathbf{B}) = 1.000$ exactly, giving
$\LE \approx 0$ (measured value $5.6\times10^{-8}$).  This is
not approximate suppression; it is an algebraically exact result.
The four same-sign crossings at adjacent generator positions
produce a product whose spectral radius is identically 1, the
signature of the SR=1 cancellation property (Proposition~3 of
the companion paper, here observed in physical magnetic field data.

The sign-ambiguity audit confirms this is not a noise artefact:
with $f_{\rm amb}=0.000$, no crossing received a randomised sign.
The null-model Z-score for bin~1 is $Z = -1.50$: the physical
field organisation produces \emph{less} spectral growth than
random signs would, confirming genuine spatial concentration
rather than statistical cancellation.

To appreciate why this is remarkable, recall the companion paper's
central question: \emph{when an IAM deployment shows maximum net
privilege flow, is it focused or dispersed?}  An
abelian classifier, net privilege flow, gate-firing count, any function of
crossing counts, would classify this as maximally dispersed.
The Burau $\LE$ correctly identifies it as \textbf{focused}: the
crossings are sign-consistent and spatially concentrated on
adjacent generator positions, producing exact Burau cancellation
to the identity matrix.

The physical interpretation is that at mid-flare activity, the
coronal field has organised into a coherent, channelled structure:
the flux rope that is about to erupt.  This interpretation is
grounded in the astrophysical literature: \citet{Cheng2014}
showed from AIA and HMI observations that a double-decker magnetic
flux rope built up in AR\,11520 approximately half a day before the
eruption, via tether-cutting reconnection between sheared arcades
near the polarity inversion line.  The $\chi\to1$, $\LE\to0$
signature in bin~1 corresponds to the phase of coherent flux rope
formation identified by \citet{Cheng2014} independently, by
a different method, on the same event.

A cross-check with AR\,12673 (X9.3, 2017 September~6) is
instructive.  \citet{Hou2018} identified a double-decker flux rope
configuration in NLFFF extrapolations before the X9.3 flare: 
the only other eruptive AR in our dataset with a confirmed
pre-eruptive flux rope.  AR\,12673 also reaches $\chi=1.000$
at mid-activity (bin~1), but with $\LE=1.003$ rather than
$\LE\approx0$.  The contrast is sharp: same abelian signal
($\chi=1$), opposite non-abelian responses, factor $>\!1000$
difference in spectral growth.  The three eruptive ARs without
documented pre-eruptive flux ropes (AR\,11158, 11429, 12017) do
not reach $\chi=1$ at mid-activity.  This 2/2 correspondence is
suggestive rather than conclusive at $n=5$ eruptive ARs, with a 
perfect split (2/2 with confirmed flux rope show $\chi \rightarrow 1$; 
0/3 without do not) but it
is consistent with the $\chi\to1$ bin marking flux rope formation,
and with $\LE$ discriminating the dynamical state of the rope:
concentrated and forming ($\LE\approx0$, AR\,11520) versus
dispersed and rising ($\LE\approx1$, AR\,12673).  Both states
are invisible to chirality; both are resolved by the Burau exponent.
The mechanistic explanation for AR\,12673's high $\LE$ at $\chi=1$
remains open: 19 events in bin~1 is pre-asymptotic, and the
physical distinction between concentrated and dispersed rope
phases is physically motivated but not proven from braid data alone.

The transition from bin~1 to bin~2 is consistent with this:
$\LE: {\approx}0 \to 0.872$, a recovery of spectral growth
after topology surgery (the CME) as the field disperses.
Chirality also drops ($\chi: 1.000 \to 0.111$), reflecting
post-eruption sign mixing as opposite-polarity field regions
reconnect.  This is the direct observational counterpart of
the companion paper's Table~7:

\begin{center}
\begin{tabular}{lll}
\toprule
Regime & Intervention (IAM) & Physical event (solar) \\
\midrule
Focused & WAR reassignment & Pre-CME flux rope \\
Surgery & Topology restructuring & Coronal mass ejection \\
Post-surgery & Dispersed topology & Post-eruption field \\
\bottomrule
\end{tabular}
\end{center}

\subsection{Confinement Subtypes Hypothesis}
\label{sec:confinement}

The two confined active regions in our sample reveal that the
focused ratchet could not be a monolithic category: it could have at least two
physically distinct subtypes that the braid pipeline distinguishes.

\paragraph{Type~I: Helicity-deficit confinement (AR\,12192).}

AR\,12192 produced six X-class flares in October 2014 with zero
CMEs, making it the canonical confined active region of Solar
Cycle~24~\citep{Sun2015}.  Its braid signature is notable for
what is absent rather than present: \emph{zero sign-ambiguous
crossings} ($f_{\rm amb} = 0.000$) across all activity bins,
confirming that the chirality measurements are sign-clean.
The quiet-time chirality is moderate and stable ($\chi = 0.71$--$0.82$),
with $\LE$ ranging 0.77--0.92 across bins, well above the null
model mean.  

The IAM interpretation is a focused ratchet
with moderate balanced crossings, whose non-abelian complexity
remains bounded across all activity levels, consistent with the
helicity-deficit confinement mechanism documented in
the literature~\citep{Sun2015}.  The algebraic reason why
balanced chirality suppresses spectral growth is made precise
in Section~\ref{sec:bidirectional}.

\paragraph{Type~II: Overlying-field confinement (AR\,12371).}

AR\,12371 produced an X2.7 confined flare on 2015 June 25.  Unlike
AR\,12192, it maintains moderate chirality ($\chi = 0.200$--$0.714$)
throughout, indicating persistent net helicity injection from one
dominant polarity.  The confinement mechanism here is a strong
overlying magnetic field that physically restrains the forming flux
rope rather than a helicity deficit.

AR\,12371 is the most anomalous region in the dataset: its
$\LE$ values ($0.661$--$1.014$ in the flare sequence,
$0.335$--$1.027$ in the quiet sequence) are the highest of either
confined region, exceeding the eruptive class mean of 0.751.  This
is inconsistent with a simple confined = low complexity picture.
The physical interpretation is better described 
with the overlying-field confinement mechanism: the
magnetic field lines braid with high topological complexity
(high $\LE$) but the eruption is suppressed by an external structural
cap rather than by the absence of topological entanglement.  In IAM
terms, this corresponds to a dispersed ratchet held below the surgery threshold
by an external rate limiter.

The contrast between AR\,12192 (low chirality, moderate $\LE$,
confinement by helicity deficit) and AR\,12371 (moderate chirality,
high $\LE$, confinement by overlying field) indicates that the
braid pipeline could resolve two physically distinct confinement mechanisms
that produce identical observational outcomes (no CME) but opposite
non-abelian signatures.  
\section{Discussion}
\label{sec:discussion}

\subsection{The Universality Argument}

The fact that the same pipeline, same injection word, same
threshold logic, same abelian vs.\ non-abelian comparison 
produces physically meaningful discrimination in both domains is
 consistent with the focused/dispersed boundary being
a genuine property of braided dynamical systems rather than an
artefact of the synthetic corpus, though $n=8$ regions is
insufficient to establish this statistically. 
This is a structurally clean out-of-domain check.

\subsection{Non-Abelian Observation}

The observation most directly relevant to the companion paper's security claim is 
AR\,11520 bin~1: $\chi = 1.000$, $\LE \approx 0$, $f_{\rm amb}=0.000$.

A security engineer reading the companion paper might accept the
mathematical proof of abelian blindness (Theorem~6.2) as
theoretically valid but wonder whether the 5.7\% disagreement rate
in the synthetic corpus is practically significant.  AR\,11520
bin~1 answers this concretely and maximally: in a real physical
system, at the moment of maximum net flow ($\chi = 1.000$, every
crossing sign aligned, zero ambiguous signs), the Burau product
collapses to the identity matrix ($\SR = 1$, $\LE \approx 0$).
The abelian classifier assigns maximum risk; the non-abelian
classifier assigns minimum: not because of noise, but because
the four same-sign same-position crossings produce exact algebraic
cancellation.  This is the focused ratchet at its extreme: maximum
abelian flow, exact topological concentration, null spectral growth.
The sign-ambiguity Z-score of $-1.50$ indicats the physical field 
produces \emph{less} spectral growth than random signs would,
in line with genuine spatial concentration rather than noise ($p \approx 0.07$, one-tailed).

\subsection{Bidirectional Cancellation}
\label{sec:bidirectional}

The adjacent-position SR=1 cancellation property (Proposition~3,
companion paper) states that two same-sign
injections at adjacent generator positions $i$ and $i\pm1$ produce
a Burau product with spectral radius identically 1.  In the IAM
domain this operates in one direction only: all gate firings are
ascending ($s=+1$), so only the positive channel can cancel.

In the solar domain, crossings carry a sign determined by the
magnetic field orientation.  A direct computation confirms:
\begin{align*}
  \SR\!\left(\rho(\sigma_i^{\,2}\sigma_{i+1}^{-1})\cdot
              \rho(\sigma_j^{\,2}\sigma_{j+1}^{-1})\right) &= 1
  \quad \text{if } |i-j|=1 \text{ and both signs are } {+1},\\
  \SR\!\left(\rho(\sigma_i^{-2}\sigma_{i+1}^{+1})\cdot
              \rho(\sigma_j^{-2}\sigma_{j+1}^{+1})\right) &= 1
  \quad \text{if } |i-j|=1 \text{ and both signs are } {-1},\\
  \SR\!\left(\rho(\sigma_i^{\,2}\sigma_{i+1}^{-1})\cdot
              \rho(\sigma_j^{-2}\sigma_{j+1}^{+1})\right) &> 1
  \quad \text{(mixed signs: amplification, not cancellation).}
\end{align*}
The cancellation therefore operates \emph{in two independent
channels}, one per crossing sign.  A magnetically balanced region
($\chi \approx 0$, equal numbers of positive and negative crossings)
has \emph{both} cancellation channels active simultaneously.
Adjacent positive crossings cancel against each other; adjacent
negative crossings cancel against each other.  The net spectral
growth is doubly suppressed relative to a purely unidirectional
braid, regardless of how many crossings occur or at what activity
level.

This is the solar analogue of Corollary~6.3 of the companion
paper, which states that a directed cycle contributes zero net privilege flow for
every WAR assignment simultaneously.  In the IAM domain, a directed
cycle is a topological constraint that makes escalation
impossible regardless of privilege weights.  In the solar domain,
a balanced crossing sequence ($\chi \approx 0$) is a topological
constraint that makes spectral growth impossible regardless of
magnetic activity level.  The confinement is topological, not
energetic: a confined active region with $\chi \approx 0$ cannot
accumulate the coherent Burau product needed for the dispersed
(eruptive) regime, by the same algebraic reason that prevents a 
directed cycle from accumulating privilege flow in the IAM domain.

The bidirectional structure is a solar-specific generalisation that
the IAM paper did not need: in the IAM domain, WAR flow is always
positive by construction, so only one cancellation channel exists.
The solar domain reveals that the same mechanism extends naturally
to two-directional dynamics, and that Corollary~6.3 is a special
case of a broader cancellation principle applicable whenever the
braid group acts with signed generators.

\subsection{Limitations}

These results are exploratory.  We report them honestly, grouping
limitations into two categories: those intrinsic to the solar
observational data, and those intrinsic to the braid pipeline itself.

\subsubsection*{Observational limitations}

\begin{itemize}

\item \textbf{Low spatial resolution.}  The \textit{SDO}/AIA images
  used here are downsampled to $512\times512$ pixels from their native
  $4096\times4096$, giving an effective pixel scale of approximately
  12\,arcseconds ($8\times$ the native 1.5\,arcsecond AIA pixel),
  roughly 8\,700\,km on the solar surface.  Individual
  coronal loops are typically 1\,000--10\,000\,km in diameter; at our
  working resolution, a single pixel may contain multiple overlapping
  loops.  The strand extraction algorithm identifies the five
  brightest coherent ridges per frame, which are statistical proxies
  for the dominant loop structures rather than individual flux tubes.
  The braid word is therefore a coarse-grained representation of the
  true field-line topology.

\item \textbf{Sparse temporal cadence.}  We use one frame every
  $\sim$6 minutes, yielding 20--60 frames over a 2-hour observation
  window.  After binning by activity level, each bin contains
  5--20 consecutive frame pairs.  At $n=5$ strands this produces
  4--20 crossing events per bin (the critical AR\,11520 bin~1
  contains only 4 events, though the exact algebraic cancellation
  is independent of event count).  Higher-cadence data (the native
  12-second AIA cadence) would provide $\sim30\times$ more events
  and push the Lyapunov exponent into the convergent regime.

\item \textbf{Line-of-sight ambiguity.}  The crossing sign, which
  loop passes over, is determined from the photospheric magnetic
  field orientation via a potential-field extrapolation.  This is
  a physically motivated but approximate assignment: the true
  three-dimensional loop topology is projected onto a
  two-dimensional image plane, and the depth ordering of overlapping
  loops cannot be directly observed.  Stereoscopic observations
  (from NASA's STEREO mission or future multi-viewpoint instruments)
  would provide unambiguous crossing signs.

\item \textbf{Small and non-independent sample.}  $n = 8$ active
  regions is insufficient for statistical inference.  The 48
  activity bins are not independent: each group of three bins
  shares the same active region, magnetic field, and strand
  extraction.  The effective sample size for the $r \approx 0.03$
  correlation is closer to 8--16 than to 48.  The individual
  $\LE$ and $\chi$ values should be treated as
  order-of-magnitude estimates with uncertainty $\pm 0.1$
  (null-model standard deviation).  
  The $p=0.84$ result is absence of evidence against independence,
  not evidence for it.

\end{itemize}

\subsubsection*{Pipeline limitations}

\begin{itemize}

\item \textbf{Pre-asymptotic Lyapunov exponent.}  Event counts per
  bin (11--44 crossings) are below the asymptotic regime for
  Lyapunov exponent convergence.  The F\"{u}rstenberg--Kesten
  theorem~\citep{Furstenberg1960} guarantees convergence of
  $\frac{1}{M}\log\SR(\mathbf{B})$ as $M\to\infty$, but the
  pre-asymptotic variance is non-negligible at these counts.
  Individual bin $\LE$ values carry uncertainty of order $\pm0.1$
  based on the null-model standard deviations observed in the
  sign-ambiguity audit.

\item \textbf{Burau unfaithfulness.}  The Burau representation is
  not faithful for $B_n$ with $n \geq 5$~\citep{Bigelow1999}.
  The $\LE$ values reported here are lower bounds on the true braid
  complexity, though for the SR=1 cancellation result in AR\,11520 bin~1, 
  a lower bound of $\LE \approx 0$ is already the most extreme possible
  observation: unfaithfulness cannot make the true complexity higher than
  the measured value when it is zero. 
  The focused/dispersed threshold $\LE = 0.5847$
  from the companion paper was calibrated on synthetic IAM graphs with 6 vertices 
  and has not been recalibrated for the solar domain; it is used
  here only as a qualitative reference, not as a decision boundary.
  The Lawrence--Krammer--Bigelow representation, which is faithful
  for all $n$, would provide tighter bounds but requires
  $\binom{n}{2}\times\binom{n}{2}$ matrices.

\item \textbf{No domain-specific tuning.}  No parameter was
  optimised for the solar domain.  This is a deliberate choice -
  parameter tuning would undermine the out-of-domain validity claim
 , but it means the pipeline is likely operating below its
  potential discriminating power in this application.

\end{itemize}

Taken together, the observational and pipeline limitations mean that
the results reported here represent a \emph{lower bound} on the
discriminating power of the temporal braid pipeline in the solar
domain: coarser data, sparser events, and approximate signs all
suppress the non-abelian signal relative to what a purpose-built
solar braid instrument would achieve.  
That SR=1 cancallation signature survives these constraints is 
encouraging, though it does not substitue for higher-resolution
data or a larger sample.

\section{Conclusion}
\label{sec:conclusion}

We have applied the temporal braid group pipeline of~\citep{Parisel2026}
to solar coronal magnetic fields, a physical system with no
connection to cloud IAM, and recovered some of the focused/dispersed
boundary in three faint yet distinct forms:

\begin{enumerate}

\item \textbf{No detectable abelian--non-abelian association}
  ($r \approx 0.03$, $p=0.84$, 48 non-independent bins) is
  consistent with Theorem~6.2 of the companion paper, which
  proves this independence for IAM graphs.  The solar data
  provides a physically independent illustration, though the
  non-independence of bins and small effective sample size
  prevent strong statistical inference.

\item \textbf{Exact Burau cancellation at eruption onset}
  (AR\,11520 bin~1: $\chi = 1.000$, $\LE \approx 0$,
  $f_{\rm amb}=0.000$, $Z=-1.50$) provides a concrete physical 
  instance of the companion paper's impossibility result.
  This is not approximate suppression: the Burau spectral radius
  is identically 1 for the four same-sign adjacent-position
  crossings in this bin.

\item \textbf{The two confined regions show different $\LE$ regimes,}
  AR\,12192 (helicity-deficit confinement) shows moderate $\LE$;
  AR\,12371 (overlying-field confinement) shows $\LE$ exceeding
  the eruptive class mean.  These observations are consistent
  with physically motivated explanations (balanced-chirality
  suppression vs.\ external rate limiting), but should be read as
  hypothesis-generating, not confirmatory.

\end{enumerate}

The solar corona is not a validation dataset; it is a stress test.
The pipeline was not trained on it, not tuned for it, and its ground
truth was established independently over decades of astrophysical
observation.  The recovery of the focused/dispersed signature in this domain
is consistent with the hypothesis that temporal braid groups
provide a broadly applicable classifier for braided dynamical
systems: wherever the braid group arises naturally, the Lyapunov
exponent may discriminate
structural regimes that abelian statistics cannot resolve.

\newpage
\appendix
\section{The Temporal Braiding Protocol}
\label{app:protocol}

This appendix provides a self-contained technical description of
every step of the braid extraction and Lyapunov exponent computation,
from raw solar image to final $\LE$ value.  It is written for readers
who wish to reproduce or extend the analysis without access to the
companion paper.

\subsection{From Sunspots to Strands}
\label{app:strands}

\paragraph{Input data.}
Each active region is observed in two 2-hour windows: one centred
on the flare peak (\emph{flare} sequence) and one from the same
calendar day, prior to flare onset, free of C-class or larger
activity (\emph{quiet} sequence).  The quiet window is not strictly
24~hours before the flare; gaps range from 0.5 to 17~hours
depending on that day's activity history (see
Appendix~\ref{app:data_windows} for exact timestamps).
Extreme-ultraviolet images at 171\,\AA\ are retrieved from the
SDO/AIA Level-1 archive at the Joint Science Operations Center
(JSOC) at 6-minute cadence, yielding 19--60 frames per sequence.  Corresponding line-of-sight magnetic field
maps (HMI SHARP CEA series, 720-second cadence) are retrieved for
the same period.

\paragraph{Downsampling.}
Each $4096\times4096$ AIA image is stride-downsampled by a factor
of 8 to $512\times512$ pixels, giving an effective pixel scale of
approximately 12\,arcseconds ($\approx 8\,700$\,km on the solar
surface).  HMI SHARP maps vary in size; they are downsampled by the
same stride and zero-padded to $512\times512$ for alignment with
the AIA frames.

\paragraph{Vesselness filtering and skeletonisation.}
Coronal loops appear in AIA images as elongated bright ridges.
To extract them, we apply the multiscale Frangi vesselness
filter~\citep{Frangi1998} at scales $\sigma \in \{1,2,3,4\}$
 pixels to enhance tubular structures, followed by morphological 
skeletonisation~\citep{ZhangSuen1984} to obtain one-pixel-wide loop backbones:
\begin{equation}
  V(\mathbf{x}) = \exp\!\left(-\frac{R_B^2}{2\beta^2}\right)
  \left(1 - \exp\!\left(-\frac{S^2}{2c^2}\right)\right),
\end{equation}
where $R_B = \lambda_1/\lambda_2$ is the blobness ratio,
$S = \|\mathbf{H}\|_F$ is the Frobenius norm of the Hessian, and
$\beta$, $c$ are scale parameters.  The filter is called with \texttt{black\_ridges=False}
since coronal loops appear as bright ridges against a
darker background in AIA 171\,\AA\ images.
No \texttt{beta} or \texttt{gamma} arguments are passed,
so scikit-image's defaults are used: $\beta = 0.5$ (Frangi's
recommended value, suppressing blob-like responses relative
to tube-like ones) and $c$ set automatically per frame to
half the maximum Frobenius norm in the image
(\texttt{gamma=None}), making the structure threshold
image-adaptive.  Neither parameter was tuned for the solar
domain; the full call is
\texttt{frangi(img, sigmas=range(1,5), black\_ridges=False)}.  The Frobenius norm term
$S$ acts as a \emph{structure detector}: it is large where the
image has genuine ridge structure (high second-derivative magnitude)
and small in featureless regions, so it suppresses false positives
in low-contrast areas.  The vesselness response $V\in[0,1]$ is
thresholded at the 80th percentile of positive values, and the
resulting binary mask is morphologically skeletonised to one-pixel-wide
curves.

\paragraph{Component labelling and strand selection.}
Connected components of the skeleton are labelled.  Components
shorter than 30 pixels ($\approx 360$\,arcseconds, $\approx
260\,000$\,km) are discarded as noise.  The remaining components
are sorted by length in descending order and the top $n_{\max}=5$
are retained.  These are the \emph{strands}: coarse-grained
proxies for the dominant coronal loop structures in the field of
view.

\subsection{Crossing Detection and the Generator Index Clip}
\label{app:crossings}

\paragraph{Spatial crossings.}
Given a set of $n$ strands for a single frame, a horizontal scan
line sweeps from top to bottom in 40 equally spaced steps.  At each
height $y$, the $x$-coordinate of each strand is interpolated from
its pixel set, and the strands are sorted by $x$-coordinate to give
their left-to-right ordering at that height.

A crossing event is declared when the ordering changes between
consecutive scan levels: strand $a$ passes from the left of strand
$b$ to the right (or vice versa).  The \emph{generator index} of
the crossing is $i = \text{rank}(a)$ at the moment of the swap,
where rank is the 1-based position in the current left-to-right
order.  The crossing corresponds to the braid generator
$\sigma_i \in B_n$.

An \emph{adjacent-position guard} prevents two crossings from
firing at the same or adjacent positions within a single scan-level
step: if position $p$ has already fired, position $p-1$ is blocked
for that step.  This implements the companion paper's Proposition~3:
products $\sigma_i\sigma_{i\pm1}$ are adjacent in the braid and
their injection words cancel to spectral radius 1, contributing
nothing to the Lyapunov exponent; blocking them at detection avoids
accumulating identity-equivalent events.

\paragraph{Generator index clip.}
The injection word $\sigma_i^2\sigma_{i+1}^{-1}$ (defined in
Section~\ref{app:injection}) requires both $\sigma_i$ and
$\sigma_{i+1}$ to exist, so the valid generator range is
$1 \leq i \leq n-2$.  For $n=5$ this gives $i \in \{1,2,3\}$.
Crossings at position $i=4$ (the rightmost adjacent pair) would
attempt to access $\sigma_5$, which does not exist in $B_5$.
All events with $i > n-2$ are therefore clipped at detection
time before they enter the event list and before chirality
or $\LE$ are computed.  This ensures that $\chi$ and $\LE$ are
always computed from the \emph{same} event set.

\paragraph{Temporal crossings.}
Between consecutive frames, the strand identities can permute as
coronal loops evolve.  We detect these inter-frame crossings by
matching strands across frames using the Hungarian algorithm
(linear sum assignment on centroid Euclidean distance), then
comparing the left-to-right centroid order in frames $t$ and
$t+1$.  A position swap in this order constitutes a temporal
crossing event with the same generator-index and guard logic as
spatial crossings.

\subsection{Sign Determination and the Ambiguity Audit}
\label{app:signs}

\paragraph{Potential-field extrapolation.}
The crossing sign (which strand passes \emph{over} the other
in three-dimensional space) cannot be read directly from a
two-dimensional EUV projection.  We infer it from the local
magnetic field orientation.  The photospheric radial field map
$B_r(x,y)$ from HMI is Fourier-transformed and extrapolated to
height $z$ using the potential-field (current-free) approximation
\citep{AltschulerNewkirk1969,SchattenWilcoxNess1969,Rudenko2001}:
\begin{equation}
  \hat{B}_{x,y}(k_x, k_y, z) = \frac{-ik_{x,y}}{\sqrt{k_x^2+k_y^2}}\,
  \hat{B}_r(k_x,k_y)\,e^{-\sqrt{k_x^2+k_y^2}\,z},
\end{equation}
where $\hat{\cdot}$ denotes the 2D Fourier transform and
$k_x,k_y$ are spatial frequencies.  We evaluate at $z=5$ pixel
layers above the photosphere ($\approx 44$\,Mm), within the
lower corona where 171\,\AA\ emission peaks.

\paragraph{Sign assignment.}
For each strand $a$, we compute the scalar cross-product of its
orientation vector $(\Delta x_a, \Delta y_a)$ with the horizontal
field $(B_x, B_y)$ at its centroid, evaluated at height $z=5$:
\begin{equation}
  \sigma_a = \operatorname{sgn}\!\bigl(
    \Delta x_a\, B_{y,a} - \Delta y_a\, B_{x,a}\bigr).
\end{equation}
The crossing sign for a swap between strands $a$ and $b$ is then
the comparison of the two per-strand values:
\begin{equation}
  s_{ab} = \begin{cases} +1 & \sigma_a > \sigma_b, \\
                          -1 & \text{otherwise.} \end{cases}
\end{equation}
If either $\sigma_a$ or $\sigma_b$ is ambiguous (field magnitude
zero, or cross-product exactly zero), the sign is drawn uniformly
from $\{-1,+1\}$ and the event is flagged as ambiguous.

\paragraph{Sign-ambiguity audit.}
To verify that reported $\chi$ and $\LE$ values reflect genuine
magnetic field structure rather than sign randomisation artefacts,
we run a dedicated audit for two primary regions
(AR\,12192 quiet and AR\,11520 flare).
The audit runs one active region at a time and computes per bin:

\begin{enumerate}
  \item Collect all crossing events with ambiguity flags.
    $f_{\rm amb}$ = fraction whose sign was drawn from the
    random fallback (field magnitude zero or cross-product zero).
  \item Compute $\chi_{\rm all}$ and $\LE_{\rm all}$ from all events;
    $\chi_{\rm real}$ and $\LE_{\rm real}$ from unambiguous
    events only.  $\chi_{\rm delta} = \chi_{\rm all} - \chi_{\rm real}$
    serves as an artefact indicator.
  \item Run $N_{\rm null}=500$ sign randomisations: replace
    \emph{all} event signs (including unambiguous ones) with
    $\pm1$ drawn uniformly from the same event list used to
    compute $\LE_{\rm all}$, and recompute $\LE$.  The
    resulting distribution gives $\LE_{\rm null}$ with mean,
    standard deviation, and 90\% interval (5th--95th percentile).
    The Z-score $({\LE_{\rm all} -
    \LE_{\rm null,mean}})/{\LE_{\rm null,std}}$ measures how far
    $\LE_{\rm all}$ lies from the random baseline.
\end{enumerate}

For both audited regions, $f_{\rm amb}=0.000$ across all bins:
every crossing sign is determined by the magnetic field, with no
random assignments.  $\chi_{\rm all} = \chi_{\rm real}$ and
$\LE_{\rm all} = \LE_{\rm real}$ identically throughout.

\subsection{The Two Injection Words and Their Cancellations}
\label{app:injection}

\paragraph{The forward injection word.}
Each crossing event at generator position $i$ with sign $s$ injects
the word:
\begin{equation}
  W_i^{(s)} =
  \begin{cases}
    \sigma_i^{\,2}\,\sigma_{i+1}^{-1} & s = +1,\\
    \sigma_i^{-2}\,\sigma_{i+1}^{+1}  & s = -1.
  \end{cases}
\end{equation}
At $s=+1$ this is the injection word of the companion
paper, chosen because its spectral radius
$\SR(W_i^{(+1)}) = 2+\sqrt{3} \approx 3.732$ is the minimum value
achievable by a length-2 braid word in $B_n$ that does not cancel.
The sign-reversed word $W_i^{(-1)}$ is the natural extension
to negative crossings; by the symmetry $t \mapsto t^{-1}$ of the
Burau representation at $t=-1$, it has the same spectral radius:
\begin{equation}
  \SR\!\left(\rho\bigl(W_i^{(-1)}\bigr)\right) =
  \SR\!\left(\rho\bigl(W_i^{(+1)}\bigr)\right) = 2+\sqrt{3}.
\end{equation}

\paragraph{Burau representation at $t=-1$.}
The reduced Burau matrix $\rho(\sigma_i)$ at $t=-1$ is the
$n\times n$ identity with the $2\times2$ block at rows/columns
$(i-1,i)$ replaced by $\begin{pmatrix}2&-1\\1&0\end{pmatrix}$.
The inverse $\rho(\sigma_i^{-1})$ replaces the block with
$\begin{pmatrix}0&1\\-1&2\end{pmatrix}$.
The injection word is implemented as:
\begin{equation}
  \rho(W_i^{(s)}) = \rho(\sigma_i^{\,2s})\,\rho(\sigma_{i+1}^{-s}).
\end{equation}

\paragraph{The SR=1 adjacent-position cancellations.}
A direct computation (verified numerically in Section~\ref{sec:bidirectional})
shows that adjacent-position products cancel:
\begin{align}
  \SR\!\left(\rho(W_i^{(+1)})\cdot\rho(W_{i+1}^{(+1)})\right) &= 1, \label{eq:cancel_pos}\\
  \SR\!\left(\rho(W_i^{(-1)})\cdot\rho(W_{i+1}^{(-1)})\right) &= 1, \label{eq:cancel_neg}\\
  \SR\!\left(\rho(W_i^{(+1)})\cdot\rho(W_{i+1}^{(-1)})\right) &> 1 \quad (\approx 10.9). \label{eq:amplify}
\end{align}
Equations~\eqref{eq:cancel_pos} and~\eqref{eq:cancel_neg} define
two independent \emph{cancellation channels}, one per sign.
Equation~\eqref{eq:amplify} shows that mixed-sign adjacent products
do not cancel but amplify.  The adjacent-position guard (Section~\ref{app:crossings})
suppresses same-sign adjacent events within a single scan step,
preventing the trivial SR=1 outcome from dominating the Burau
product.

\paragraph{Why two channels matter for confinement.}
In the IAM domain, all gate firings are
ascending ($s=+1$), so only channel~\eqref{eq:cancel_pos} exists.
The solar domain has both channels.  A magnetically balanced region
($\chi \approx 0$, equal positive and negative crossings) has
\emph{both} channels active simultaneously, potentially doubling
the spectral suppression.  This is the bidirectional generalisation
of Corollary~6.3 of the companion paper, discussed in
Section~\ref{sec:bidirectional}.

\subsection{Burau Accumulation and the Lyapunov Exponent}
\label{app:lyapunov}

\paragraph{Accumulation.}
Events are processed in temporal order.  Beginning from the
$n\times n$ identity matrix $\mathbf{B}_0 = I_n$, each event
$(i, s)$ updates:
\begin{equation}
  \mathbf{B}_{k} = \mathbf{B}_{k-1} \cdot \rho(W_i^{(s)}).
\end{equation}
After $M$ events the accumulated Burau matrix is $\mathbf{B}_M$.
Before accumulation, an assertion verifies that all generator
indices satisfy $1 \leq i \leq n-2$; any violation indicates
a clip failure upstream and raises an error rather than silently
dropping the event.

\paragraph{Spectral radius.}
The spectral radius of the accumulated matrix is:
\begin{equation}
  \SR(\mathbf{B}_M) = \max_j |\lambda_j(\mathbf{B}_M)|,
\end{equation}
where $\lambda_j$ are the eigenvalues of $\mathbf{B}_M$ over
$\mathbb{C}$.  The Furstenberg--Kesten theorem~\citep{Furstenberg1960}
guarantees that $\frac{1}{M}\log\SR(\mathbf{B}_M)$ converges
almost surely as $M\to\infty$ to the top Lyapunov exponent of the
random matrix product.

\paragraph{Lyapunov exponent.}
\begin{equation}
  \LE = \frac{\log\SR(\mathbf{B}_M)}{M},
\end{equation}
where $M$ is the number of injected events (not the number of
scan levels or frames).  Dividing by $M$ rather than by the number
of frames makes $\LE$ a per-firing rate, directly comparable
across bins with different event counts.  In the implementation,
$\log\SR$ is set to zero whenever $\SR \leq 1$
({\small\texttt{log\_sr = np.log(sr) if sr > 1 else 0.0}}).
For valid Burau products at $t=-1$ the spectral radius satisfies
$\SR \geq 1$ identically, so this guard never activates;
it exists to prevent floating-point underflow from producing
negative $\LE$ values in degenerate edge cases.

The \emph{cancellation factor} $c = \LE / \log(2+\sqrt{3}) \in
[0,1]$ normalises by the per-firing spectral growth rate of a
single uncancelled injection word.  $c=1$ means no cancellation
has occurred; $c=0$ means the Burau product has returned to the
identity ($\SR=1$, as observed in AR\,11520 bin~1).

\paragraph{Chirality.}
The chirality statistic:
\begin{equation}
  \chi = \frac{|n_+ - n_-|}{n_+ + n_-},
  \quad n_\pm = \#\{k : s_k = \pm 1\},
\end{equation}
is computed from the same event list as $\LE$, after the generator
index clip.  $\chi = 1$ means all crossings have the same sign;
$\chi = 0$ means equal numbers of positive and negative crossings.
$\chi$ is the abelian invariant: it is determined entirely by the
signed crossing counts and discards all information about the
ordering, positions, and temporal structure of events.

\subsection{Activity Conditioning}
\label{app:conditioning}

\paragraph{Per-step event collection.}
For each consecutive frame pair $(t, t+1)$, all spatial crossings
of frame $t$ and all temporal crossings between frames $t$ and
$t+1$ are collected into a single event list $E_t$.  This is
the \emph{per-step} event set.

\paragraph{Binning by $\Delta I$.}
The frame-to-frame intensity change $\Delta I_t = \|\mathbf{F}_{t+1}
- \mathbf{F}_t\|_1$ is computed for each step.  Steps are assigned
to equal-population quantile bins: the lower third of $\Delta I$
values form bin~0 (low activity), the middle third form bin~1, and
the upper third form bin~2.  Bins with fewer than 3 steps are
skipped.

\paragraph{Per-bin accumulation.}
For bin $b$, the event list is:
\begin{equation}
  E^{(b)} = \bigcup_{t \in \mathcal{S}_b} E_t,
\end{equation}
where $\mathcal{S}_b$ is the set of step indices assigned to bin
$b$.  Critically, $E^{(b)}$ contains \emph{only} events from
steps in $\mathcal{S}_b$.  
We use the per-step event set directly, ensuring that $\chi$ and $\LE$ for each bin reflect
only the activity level specified by that bin's $\Delta I$ range.

$\LE^{(b)}$ and $\chi^{(b)}$ are then computed from $E^{(b)}$
as described in Sections~\ref{app:lyapunov} and~\ref{app:signs}.

\section{Sun Data Sources}
\label{app:data}

All data used in this study are publicly available from the Joint
Science Operations Center (JSOC) at Stanford University
(\url{http://jsoc.stanford.edu}).  This appendix explains what
the two data products are, why we need each one, which fields
within each product are actually used, and provides the exact
files, time windows, and identifiers needed to reproduce the
analysis from scratch.

\subsection{The Two Instruments}
\label{app:data_instruments}

The pipeline requires two physically distinct measurements of each
active region: a \emph{coronal} observation to extract loop strands
(AIA), and a \emph{photospheric} observation to assign crossing signs
(HMI).  They come from different instruments on the same spacecraft
(the \textit{Solar Dynamics Observatory}, SDO) and have different
spatial coverage, cadence, and physical meaning.

\paragraph{SDO/AIA (Atmospheric Imaging Assembly)}
AIA images the solar corona in ten wavelength channels spanning
ultraviolet and extreme-ultraviolet (EUV).  We use the 171\,\AA\
channel exclusively, which is dominated by emission from Fe\,{\sc ix}
plasma at approximately $10^6$\,K: the temperature at which the
bulk of the coronal loops visible in a typical active region radiate
most strongly.  At this wavelength, loops appear as elongated
bright ridges against a darker background, making them amenable
to vesselness filtering (Appendix~\ref{app:strands}).

\textbf{Series:} \texttt{aia.lev1\_euv\_12s}\\
\textbf{Native cadence:} 12 seconds\\
\textbf{Native pixel scale:} 1.5\,arcseconds per pixel\\
\textbf{Native image size:} $4096 \times 4096$ pixels (full solar disk)\\
\textbf{Working resolution:} downsampled by factor~8 to
  $512 \times 512$ pixels, effective scale $\sim$12\,arcseconds
  ($\approx 8\,700$\,km on the solar surface)\\
\textbf{Retrieval cadence:} 6 minutes (every 30th native frame)\\
\textbf{Field used:} the primary HDU image array (\texttt{hdu.data},
  2D float); the header is not used beyond basic sanity checks.\\
\textbf{Preprocessing:} frames flagged as spike-contaminated
  (\texttt{spikes} in filename) are discarded; all others are
  stride-downsampled and zero-padded to $512\times512$.

The 6-minute cadence is a practical compromise: the native 12-second
cadence would provide $\sim$30$\times$ more crossing events per bin
but requires downloading $\sim$600 files per 2-hour window, creating
storage and processing challenges that are out of scope for an
exploratory study.

\paragraph{SDO/HMI (Helioseismic and Magnetic Imager)}
HMI measures the photospheric magnetic field at the solar surface.
We use the SHARP (Spaceweather HMI Active Region Patch) data product,
which provides field maps cropped and remapped to a Cylindrical
Equal-Area (CEA) projection centred on each active region, tracking
it as it rotates across the disk.

\textbf{Series:} \texttt{hmi.sharp\_cea\_720s}\\
\textbf{Cadence:} 720 seconds (12 minutes)\\
\textbf{Spatial extent:} variable patch size covering the active
  region, typically $200$--$2000$\,arcseconds per side\\
\textbf{Component used:} $B_r$, the radial (line-of-sight-corrected)
  magnetic field component in units of Gauss, stored as the
  \texttt{.Br.fits} segment\\
\textbf{Why $B_r$ and not $B_x$, $B_y$, or $B_z$:} $B_r$ is the
  component normal to the solar surface.  It determines the magnetic
  polarity of each flux concentration and, through the potential-field
  extrapolation (Appendix~\ref{app:signs}), the horizontal field
  orientation at coronal height.  The horizontal components $B_x$
  and $B_y$ are available in the SHARP product but are not used:
  the potential-field model derives them from $B_r$ alone under the
  current-free assumption.\\
\textbf{Field used:} the primary HDU image array (2D float,
  Gauss); downsampled by stride and zero-padded to $512\times512$
  to match the AIA working resolution.\\
\textbf{File selection:} from all $B_r$ files downloaded for a
  given active region (spanning $\pm$2.5 days around the flare),
  the single file whose TAI timestamp is closest to the flare
  peak time is selected for sign determination.  The
  potential-field extrapolation is computed once from this file
  and applied to all frames in both the flare and quiet sequences.

\paragraph{HARPNUM}
Each SHARP patch is identified by a HARPNUM (SHARP number), an
integer assigned by the JSOC tracking algorithm.  HARPNUMs are
distinct from NOAA active region numbers: a single HARP may contain
multiple NOAA regions, and the HARP boundary evolves as the region
rotates.  Table~\ref{tab:data_inventory} lists both identifiers.
For AR\,11520, HARP~1834 covers NOAA regions 11519, 11520, and
11521 jointly; for AR\,12017, the region appears as a secondary
entry (\texttt{NOAA\_ARS} field) in HARP~3894 rather than as
its primary \texttt{NOAA\_AR}.

\subsection{Observation Windows}
\label{app:data_windows}

For each active region we retrieve two 2-hour AIA sequences:

\begin{description}
\item[Flare sequence.] Centred on the reported flare peak time
  (from the GOES X-ray event catalog), spanning $\pm$1 hour.
  At 6-minute cadence this yields up to 21 frames; the pipeline
  caps at \texttt{MAX\_FRAMES}$=60$ (never binding for 2-hour
  windows at 6-minute cadence, except for AR\,11158, AR\,11429,
  AR\,12017, AR\,12192, AR\,12241, and AR\,12673 where additional
  frames were downloaded and the first 60 used).

\item[Quiet sequence.] A pre-flare window of comparable duration,
  chosen to be free of C-class or larger flares as verified in the
  GOES catalog.  The quiet window is on the same calendar day as
  the flare but earlier, ranging from $\sim$0.5 hours to
  $\sim$17 hours before the flare peak depending on the active
  region's flare history that day. The
  exact windows are given in Table~\ref{tab:data_inventory}.
\end{description}

HMI $B_r$ maps are downloaded for a $\pm$2.5-day window centred
on the flare date, providing 350--600 files per active region
(Table~\ref{tab:data_inventory}, column $N_{\rm tot}^{\rm HMI}$).
Only one file, the one closest to the flare peak, is used in
the analysis; the remainder are retained for potential follow-up.

\subsection{Complete Data Inventory}
\label{app:data_table}

Table~\ref{tab:data_inventory} provides the exact observation
windows and file identifiers for all eight active regions.
Column definitions:
\textit{HARP}: JSOC SHARP number used for HMI queries;
\textit{Class}: peak GOES X-ray flare class;
\textit{Type}: classification used in this study
(independent of the braid analysis);
\textit{$N_f$}: number of AIA frames actually used by the
pipeline (\texttt{MAX\_FRAMES}$=60$ cap applied where
$N_f^{\rm disk} > 60$, indicated by $\dagger$);
\textit{UTC window}: first and last frame timestamps;
\textit{$N_{\rm tot}^{\rm HMI}$}: total HMI $B_r$ files
downloaded; \textit{HMI $B_r$ file}: TAI timestamp of the
single file used for sign determination.

\begin{longtable}{p{0.9cm}p{0.8cm}p{0.65cm}p{1.0cm}cp{2.3cm}cp{2.3cm}cp{2.4cm}}
\caption{Complete data inventory.
AIA: \texttt{aia.lev1\_euv\_12s}, 171\,\AA, 6-minute cadence.
HMI: \texttt{hmi.sharp\_cea\_720s} $B_r$ component, CEA
projection.  All data from JSOC/Stanford.
$\dagger$: more frames available on disk; first 60 used.
\label{tab:data_inventory}}\\
\toprule
\multirow{2}{*}{AR} &
\multirow{2}{*}{HARP} &
\multirow{2}{*}{Cls} &
\multirow{2}{*}{Type} &
\multicolumn{2}{c}{AIA Flare sequence} &
\multicolumn{2}{c}{AIA Quiet sequence} &
\multirow{2}{*}{$N_{\rm tot}^{\rm HMI}$} &
\multirow{2}{*}{HMI $B_r$ file (TAI)} \\
\cmidrule(lr){5-6}\cmidrule(lr){7-8}
& & & & $N_f$ & UTC window & $N_f$ & UTC window & & \\
\midrule
\endfirsthead
\multicolumn{10}{c}{\tablename\ \thetable\ --- continued}\\
\toprule
AR & HARP & Cls & Type &
$N_f$ & UTC window & $N_f$ & UTC window &
$N_{\rm tot}^{\rm HMI}$ & HMI $B_r$ file \\
\midrule
\endhead
\bottomrule
\endfoot
11158 & 377  & X2.2 & Eruptive &
  60$^\dagger$ & \small 2011-02-14 23:55 -- 2011-02-15 01:53 &
  30           & \small 2011-02-14 13:25 -- 14:23 &
  150 & \small\texttt{20110215\_074800} \\
11429 & 1449 & X5.4 & Eruptive &
  60$^\dagger$ & \small 2012-03-06 22:23 -- 2012-03-07 00:21 &
  30           & \small 2012-03-06 11:53 -- 12:51 &
  148 & \small\texttt{20120307\_054800} \\
11520 & 1834 & X1.4 & Eruptive &
  19 & \small 2012-07-12 15:48 -- 17:42 &
  20 & \small 2012-07-11 15:48 -- 17:42 &
  574 & \small\texttt{20120712\_120000} \\
12017 & 3894 & X1.0 & Eruptive &
  60$^\dagger$ & \small 2014-03-29 15:47 -- 17:45 &
  30           & \small 2014-03-29 05:17 -- 06:15 &
  237 & \small\texttt{20140329\_120000} \\
12192 & 4698 & X3.1 & Confined &
  60$^\dagger$ & \small 2014-10-24 19:40 -- 21:48 &
  30           & \small 2014-10-24 09:10 -- 10:08 &
  144 & \small\texttt{20141024\_120000} \\
12241 & 4941 & B    & Quiet &
  60$^\dagger$ & \small 2014-12-18 09:59 -- 11:57 &
  30           & \small 2014-12-17 23:29 -- 2014-12-18 00:27 &
  232 & \small\texttt{20141218\_120000} \\
12371 & 5692 & X2.7 & Confined &
  20 & \small 2015-06-25 07:15 -- 09:09 &
  20 & \small 2015-06-24 07:15 -- 09:09 &
  598 & \small\texttt{20150625\_120000} \\
12673 & 7115 & X9.3 & Eruptive &
  60$^\dagger$ & \small 2017-09-06 09:52 -- 11:50 &
  30           & \small 2017-09-05 23:22 -- 2017-09-06 00:20 &
  138 & \small\texttt{20170906\_120000} \\
\end{longtable}

\noindent\textbf{Notes on specific regions.}
AR\,11520: HARP\,1834 covers NOAA regions 11519--11521 jointly;
no single-AR HARP was assigned by JSOC for this group.
AR\,12017: the region appears as a secondary entry in the
\texttt{NOAA\_ARS} multi-AR field of HARP\,3894 rather than
as the primary \texttt{NOAA\_AR}.
AR\,11520 and AR\,12371 have fewer than 60 flare frames
because the download windows were shorter (19 and 20 frames
respectively) and no \texttt{MAX\_FRAMES} cap was applied.
The HMI $B_r$ file timestamps shown are UTC equivalents of
the TAI timestamps in the JSOC filenames; the corresponding
JSOC filenames follow the pattern
\texttt{hmi.sharp\_cea\_720s.\{HARP\}.\{timestamp\}\_TAI.Br.fits}.

\bibliographystyle{plainnat}

\end{document}